\documentclass[prd,aps,reprint,noeprint,superscriptaddress,amsmath, amssymb]{revtex4-1}
\usepackage{graphicx}
\usepackage{subfigure}
\usepackage{lineno}
\usepackage{epsfig}
\usepackage{multirow}
\usepackage{overpic}

\usepackage{color}
\usepackage{bm} 
\usepackage{xspace} 
\usepackage{dcolumn}
\usepackage{booktabs,tabularx}
\usepackage{array}
\usepackage{textcomp}
\usepackage[colorlinks,linkcolor=blue,urlcolor=blue,citecolor=blue]{hyperref}
\usepackage{tikz}
\usepackage{mathrsfs}

\newcommand{\BESIIIorcid}[1]{\href{https://orcid.org/#1}{\hspace*{0.1em}\raisebox{-0.45ex}{\includegraphics[width=1em]{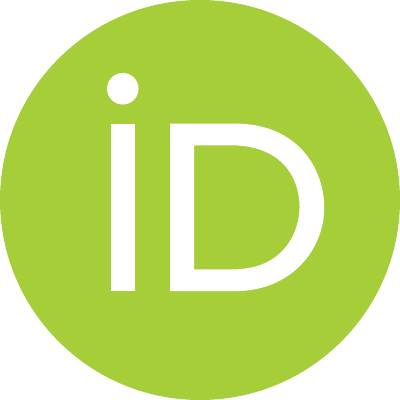}}}}

\usepackage{cleveref}

\newcommand
{\EE}{e^+e^-}

\newcommand{\XsXs}{\Xi(1530)^0\bar\Xi(1530)^0}
\newcommand{\Xs}{\Xi(1530)^0}
\newcommand{\Xsb}{\bar\Xi(1530)^0}

\begin{document}
\normalsize
\parskip=5pt plus 1pt minus 1pt


\title{
\boldmath Observation of $\Xs$ Polarization and Determination of Its Electric and Magnetic Moments in $\psi(3686)\to\XsXs$
}
\author{BESIII Collaboration}
\thanks{Full author list given at the end of the Letter.}

\begin{abstract}
Using the data sample of $2.7\times10^9$ $\psi(3686)$ events collected with the BESIII detector at the BEPCII collider, we present an observation of the $\Xi(1530)^0$ polarization in the decay $\psi(3686)\to\XsXs$ with a significance larger than $20\sigma$ compared with all other tested hypotheses. The helicity amplitudes for the process $\psi(3686)\to\XsXs$ and the moduli of form factors including electric charge, magnetic dipole, electric quadrupole, and magnetic octupole are measured for the first time by performing an angular distribution analysis. Additionally, the polarization correlations between $\Xs$ and $\Xsb$ are measured.
\end{abstract}

\maketitle


Currently, the widely accepted theory of the strong interaction is the quantum chromodynamics (QCD), a non-Abelian gauge field theory that describes the interactions between quarks and gluons, featuring quark asymptotic freedom and confinement. Since its inception over five decades ago, QCD has been precisely tested at high energy scales.
At low energy scales, in the exploration of baryons and their excited states, lattice QCD has played a pivotal role. By providing a nonperturbative computational framework, lattice QCD enables the direct calculation of hadron form factors, including the higher-order multipole moments that characterize high-spin particles~\cite{Nozawa:1990gt,Alexandrou:2008bn,Alexandrou:2010jv}. These multipole moments, i.e., form factors, significantly advance our understanding of baryon structure and the underlying dynamics of strong interactions.
Despite the successes obtained so far, the nonperturbative aspects of QCD at low energy scales remain poorly understood, posing a significant challenge to the field. More recently, the development of the angular distribution analysis theory within the helicity formalism, combined with the experimental conditions for baryon pair production in $\EE$ collisions, has enabled high-spin baryons to serve as new probes for precisely testing QCD~\cite{Alexandrou:2008bn,Alexandrou:2010jv}. This is supported by the direct experimental measurement of their higher-order multipole moments, which were previously inaccessible.

Baryons, the simplest colorless systems formed by the three-color quark combination in QCD, are among the most important probes in particle and nuclear physics for understanding quark-gluon structures. However, their size~\cite{Pohl:2010zza,Carlson:2015jba}, spin~\cite{Aidala:2012mv}, and intrinsic structure~\cite{Miller:2007uy} are still not fully understood.
Since the first experimental determination of proton form factors in the 1950s~\cite{Punjabi:2015bba}, form factors have remained a fundamental property of baryons and have been extensively studied to this day.
In recent years, baryon pair production in $\EE$ collisions has provided an ideal experimental framework for extracting timelike form factors. These timelike form factors are intimately connected with more intuitive spacelike quantities, such as charge and magnetization densities, via dispersion relations~\cite{Belushkin:2006qa,Yan:2023yff}.
According to the optical theorem, the form factors in the spacelike region are real at the lowest order due to the hermiticity of the electromagnetic Hamiltonian, while in the timelike region they are complex, leading to a transverse polarization for the produced baryons~\cite{Dubnickova:1992ii,Tomasi-Gustafsson:2005svz,Denig:2012by}.
Based on this approach, BESIII has conducted extensive measurements of baryon form factors, employing angular distribution analysis~\cite{Wang:2018kdh,BESIII:2020lkm,BESIII:2021cvv,Wang:2022zyc, BESIII:2021ypr,BESIII:2022lsz,Liu:2023xhg, BESIII:2023euh,BESIII:2023ynq,BESIII:2023lkg,BESIII:2023cvk,BESIII:2024dmr,BESIII:2024nif,BESIII:2025yzk,BESIII:2025dke,BESIII:2026hgj}. 
However, most of these studies have focused on spin-1/2 ground state baryons, while research on higher-spin baryons remains scarce, with only a limited number of investigations targeting the $\Omega^-$ baryon~\cite{Dobbs:2014ifa,Dobbs:2017hyd,BESIII:2020lkm,BESIII:2022kzc, Zhang:2026qjt, Zhang:2025qmo}.

In contrast to spin-1/2 baryons, which are described by two form factors for their charge and magnetic moment, spin-3/2 baryons are characterized by four form factors, representing their electric charge($|G_E|$), magnetic dipole ($|G_M|$), electric quadrupole ($|G_Q|$), and magnetic octupole ($|G_O|$)~\cite{Korner:1976hv}, providing a broader physical insight into the baryon structure and the properties of QCD at low energy scales. Their polarization is also more complex, involving fifteen independent components~\cite{Song:1967vpt,Doncel:1972ez,Zhao:2022lbw,Zhang:2023wmd}.
A previous BESIII study on the $\Omega^-$ has observed clear multipole polarization effects~\cite{BESIII:2020lkm}, but the single-tagging method has limited the full characterization of polarization in all directions. Moreover, investigations of baryon excited states are still lacking.
As a member of the SU(3) decuplet, $\Xs$ is a well-understood cascade resonance, providing a solid foundation for experimental analysis. The angular distribution analysis based on the double-tagging method~\cite{Zhang:2023box}, combined with the multistage cascade decay of $\Xs$, further provides favorable conditions and robust constraints for clarifying the spin polarization and correlations in all directions between the baryon and the antibaryon.

In this Letter, based on $(2712.4\pm14.3)\times10^6$ $\psi(3686)$ events~\cite{BESIII:2024lks} collected with the BESIII detector~\cite{BESIII:2009fln} at the BEPCII collider~\cite{Yu:2016cof}, we present the first multidimensional angular distribution analysis for the process $\EE\to\psi(3686)\to\Xs\Xsb$ and its subsequent decays. As a result, we report the observation and correlation of multipole polarization and helicity parameters of $\Xs$, along with the moduli of its form factors and their ratios for the first time.


To maximize reconstruction efficiency while minimizing background contamination, a double-tagging reconstruction technique is applied to the decays $\Xs\to\Xi^-\pi^+$, $\Xi^-\to\Lambda\pi^-$, $\Lambda\to p\pi^-$ and $\Xsb\to\bar\Xi^+\pi^-$, $\bar\Xi^+\to\bar\Lambda\pi^+$, $\bar\Lambda\to \bar{p}\pi^+$. Each signal event contains eight charged final state particles.
Monte Carlo (MC) simulated data samples produced with a {\sc geant4}-based~\cite{G42002iii} software package, which includes the geometric description of the BESIII detector and the detector response, are used to determine detection efficiencies and to estimate backgrounds. The simulation models the beam energy spread and initial state radiation (ISR) in the $e^+e^-$ annihilations with the generator {\sc kkmc}~\cite{Jadach2001}. The inclusive MC sample includes the production of the $\psi(3686)$ resonance, the ISR production of the $J/\psi$, and the continuum processes incorporated in {\sc kkmc}~\cite{Jadach2001}.
The exclusive MC samples for the signal process are generated using a phase space (PHSP) model implemented in the EvtGen event generator~\cite{Lange:2001uf,Ping:2008zz}.

Charged tracks must be reconstructed in the main drift chamber (MDC) within the angular region $|\cos\theta|<0.93$. Here, $\theta$ is the polar angle relative to the $z$ axis in the laboratory frame, which is the symmetry axis of the MDC. 
The number of positive or negative charged tracks must be at least three for each case.
Particle identification for charged tracks combines the measurements of the energy deposited in the MDC (${\rm d}E/{\rm d}x$) and the time of flight to form likelihoods $\mathcal{L}(h)$ $(h=p,K,\pi)$. 
Tracks are identified as protons when the proton hypothesis has the greatest likelihood $[\mathcal{L}(p)>\mathcal{L}(K)~{\rm and}~\mathcal{L}(p)>\mathcal{L}(\pi)]$, while charged pions are identified by requiring $[\mathcal{L}(\pi)>\mathcal{L}(K)~{\rm and}~\mathcal{L}(\pi)>\mathcal{L}(p)]$. 
Events with one proton, one antiproton, at least three pions of the same charge, and at least two pions of the opposite charge are selected for the analysis.
For the pion type ($\pi^+$ or $ \pi^-$) with only two candidates, one additional missing pion of the same charge is assumed in the reconstruction.
In cases where both the numbers of $\pi^+$ and $\pi^-$ candidates exceed two, the reconstruction procedure is performed by treating either one $\pi^+$ or one $\pi^-$ candidate as a missing particle, the motivation for which will be clarified later.
The combination yielding the smallest total $\chi^2$, defined as the sum of the contributions from the secondary vertex fit and kinematic fit described below, is selected.
Given the identical reconstruction logic for both cases, the following description focuses solely on the missing $\pi^-$ case. 

\begin{figure}[!htbp]
    \begin{center}
    \includegraphics[width=0.4\textwidth]{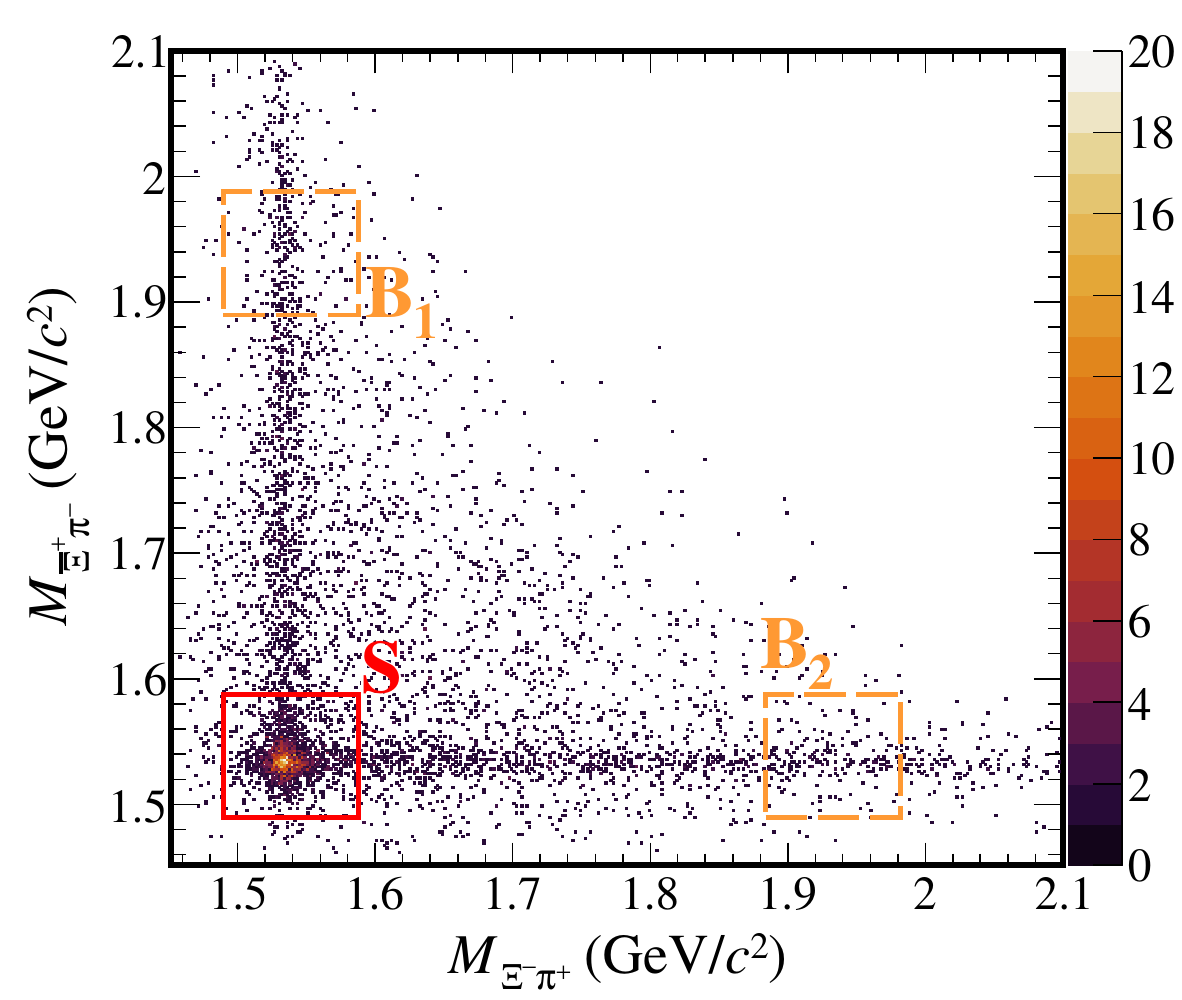}
    \end{center}
    \vspace{-0.5cm}
    \caption{Distribution of $M_{\bar\Xi^+\pi^-}$ versus $M_{\Xi^-\pi^+}$. The solid box marked by S shows the signal region, and the dashed boxes marked by $B_1$ and $B_2$ denote the sideband regions.}
    \label{2Dbox}
\end{figure}

To reconstruct $\Lambda(\bar\Lambda)$, $\Xi^-(\bar\Xi^+)$, and $\Xs$ candidates, all combinations of final state particle candidates are created, and vertex fit and secondary vertex fit~\cite{Xu:2009zzg} are applied sequentially.
The combination with the minimum quadratic sum of the residuals between the reconstructed invariant masses of all intermediate states and their corresponding known values~\cite{ParticleDataGroup:2024cfk} is selected.
Additionally, the $\Lambda(\bar\Lambda)$ and $\Xi^-(\bar\Xi^+)$ decay lengths  are required to be greater than zero to suppress backgrounds from non-$\Lambda(\bar\Lambda)$ and non-$\Xi^-(\bar\Xi^+)$ events, where negative values of decay length can result from detector resolution.
To further suppress combinatorial backgrounds and improve detector resolution, a one-constraint ($1C$) kinematic fit is employed. The fit ensures energy–momentum conservation with four constraints, while the missing $\pi^-$ track from $\Xsb$ is treated as 3 additional degrees of freedom, and events are required to satisfy $\chi^2_{1C} < 40$.
Note that even when sufficient $\pi^-$ candidates are available, full reconstruction does not reduce the background level because the backgrounds have the same final state as discussed below. Therefore, it is also treated using the missing particle approach to improve efficiency.
To eliminate the primary background from the $\EE\to\pi^+\pi^-J/\psi\to\pi^+\pi^-\Xi^-\bar\Xi^+$ process, a veto requirement of $M^{\rm Recoil}_{\pi^+\pi^-}\notin[3.077,~3.117]$ GeV/$c^2$ is applied, where $M^{\rm Recoil}_{\pi^+\pi^-}$ is the recoil mass of the $\pi^+\pi^-$ combination.
After applying the above selection criteria, the invariant masses of the $p\pi^-(\bar{p}\pi^+)$, $\Lambda\pi^-(\bar\Lambda\pi^+)$, and $\Xi^-\pi^+(\bar\Xi^+\pi^-)$ combinations are required within the ranges [1.111, 1.121], [1.314, 1.330], and [1.490, 1.588] GeV/$c^2$, respectively.
Figure~\ref{2Dbox} shows the two-dimensional distribution of $M_{\bar\Xi^+\pi^-}$ versus $M_{\Xi^-\pi^+}$, where $M_{\bar\Xi^+\pi^-}$ and $M_{\Xi^-\pi^+}$ represent the invariant masses of the $\bar\Xi^+\pi^-$ and $\Xi^-\pi^+$ pairs, respectively. A clear enhancement within the mass requirement for the $\Xi^-\pi^+(\bar\Xi^+\pi^-)$ combination, marked by the red box, is observed.

An inclusive MC sample of $\psi(3686)$, matching the size of the data, is used to study the background sources. The primary background in this analysis arises from the two three-body processes, $\psi(3686)\to \bar\Xi^+\pi^-\Xi(1530)^0$ and $\psi(3686)\to \Xi^-\pi^+\bar\Xi(1530)^0$, which exhibit peaks in the distribution of $M_{\Xi^-\pi^+}$ and $M_{\bar\Xi^+\pi^-}$ on the side associated with the excited $\Xi$ states. 
The impact of other nonpeaking backgrounds, mainly $\psi(3686)\to\pi^+\pi^-J/\psi$ and $\psi(3686)\to\pi^+\pi^-\Xi^-\bar\Xi^+$, is evaluated via MC simulations, yielding 2.2\% and 0.9\%, respectively.
The contribution of peaking backgrounds is estimated from sideband regions, chosen with reference to the distributions from exclusive MC simulations, and defined as 
$B_1$, $M_{\Xi^-\pi^+} \in [1.490,~1.588]$~GeV/$c^2$ $\cap$ $M_{\bar\Xi^+\pi^-} \in [1.890,~1.988]$~GeV/$c^2$, and $B_2$, $M_{\Xi^-\pi^+} \in [1.884,~1.982]$~GeV/$c^2$ $\cap$ $M_{\bar\Xi^+\pi^-} \in [1.490,~1.588]$~GeV/$c^2$, each with the same size as the signal region. These regions are marked with the dashed boxes in Fig.~\ref{2Dbox}.
Based on the sideband method, the signal yield is extracted as $N_{\rm sig} = N_{\rm obs} - N_{\rm bkg} = 1894\pm53$, where $N_{\rm obs}$ is the number of events in the signal region, and $N_{\rm bkg}=451\pm21$ represents the estimated background yield calculated as $N_{B_1} + N_{B_2}$. Here, $N_{B_1}$ and $N_{B_2}$ denote the numbers of events in the sideband regions, respectively.

\begin{figure}[!hbpt]
	\begin{center}
        \includegraphics[width=0.48\textwidth]{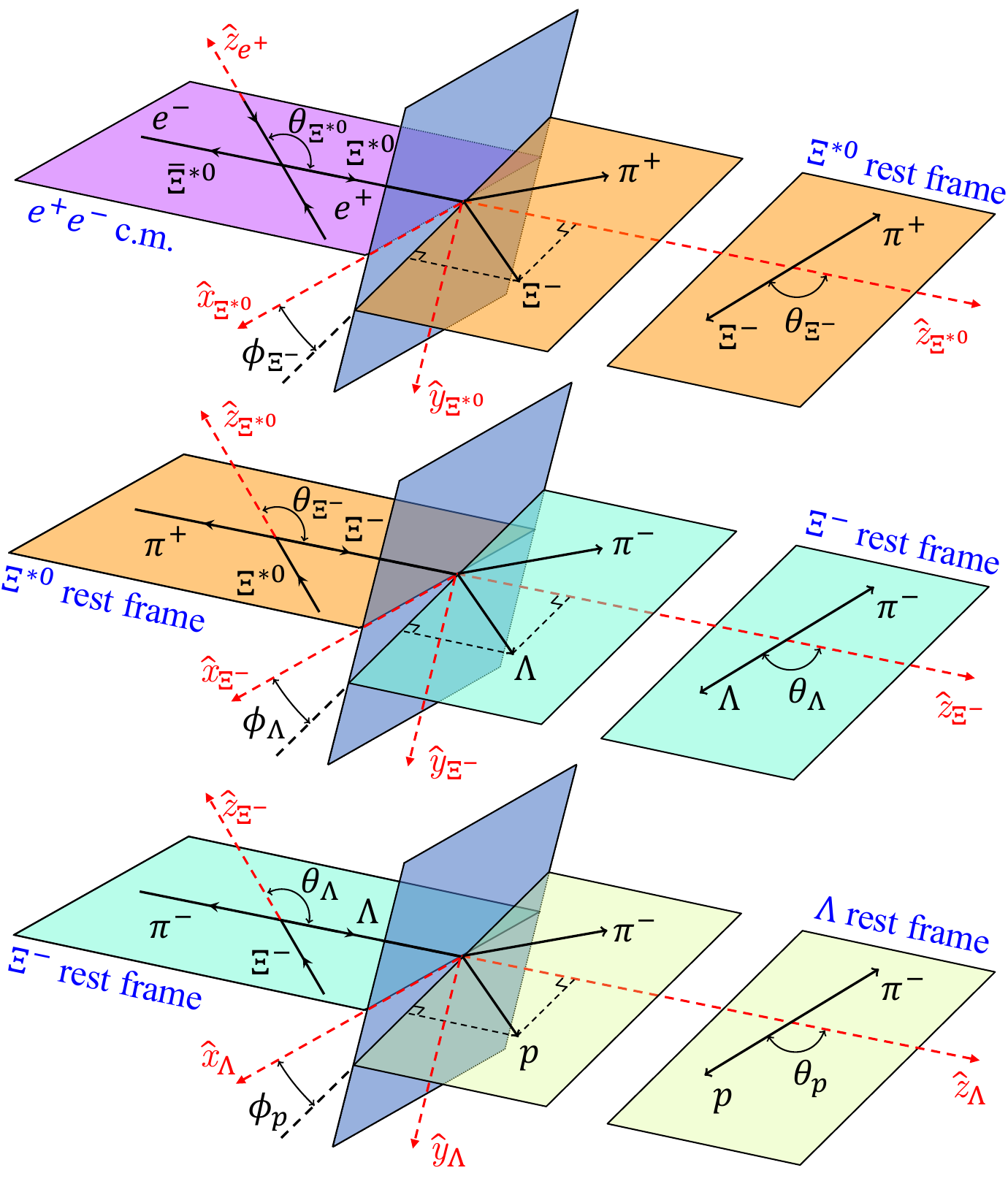}
	\end{center}
        \vspace{-0.5cm}
	\caption{The definitions of the helicity angles. The $\Xi^{*0}$ and $\bar\Xi^{*0}$ denote $\Xs$ and $\Xsb$, respectively. The polar angle $\theta_{\Xi^{*0}}$ is the angle between the $\Xs$ momentum and the $e^+$ beam direction in the $\EE$ center-of-mass frame, where the $\hat{z}_{e^+}$ axis is defined along the $e^+$ momentum. $\theta_{\Xi^-}$ and $\phi_{\Xi^-}$ are the polar and azimuthal angles of the $\Xi^-$ momentum direction in the $\Xs$ rest frame, where $\hat{z}_{\Xi^{*0}}$ is defined along the $\Xs$ momentum direction in the $\EE$ center-of-mass frame, and $\hat{y}_{\Xi^{*0}}$ is defined by $\hat{z}\times\hat{z}_{\Xi^{*0}}$. The definitions of $\theta_{\Lambda}$, $\phi_{\Lambda}$, $\theta_{p}$, and $\phi_{p}$ follow the same convention as above.}
	\label{xyz}
\end{figure}

The joint angular distribution describes the spin correlation and polarization of $\Xs$ and $\Xsb$, which is expressed as~\cite{Jacob:1959at,Perotti:2018wxm,BESIII:2020lkm,Zhang:2023box}
\begin{align}
    \mathscr{W}(\bm{\xi}; \bm{\Omega}) ={}& \sum_{\mu=0}^{15}\sum_{\nu=0}^{15}\sum_{m=0}^{3}\sum_{n=0}^{3}\sum_{i=0}^{3}\sum_{j=0}^{3} S_{\mu\nu} b_{\mu m}^{\Xi(1530)^0} b_{\nu n}^{\bar\Xi(1530)^0} \notag \\
    & \times a_{mi}^{\Xi^-} a_{nj}^{\bar\Xi^+} a_{i0}^{\Lambda} a_{j0}^{\bar\Lambda}.
    \label{cs}
\end{align}
Here, the matrix $S$ represents the spin density matrix, which is formulated in terms of four helicity amplitudes~\cite{Perotti:2018wxm}, 
while $b$ and $a$ are the polarization transfer matrices for spin-3/2 and spin-1/2 particles, respectively. The vector $\bm{\xi}$ represents thirteen helicity angles, as illustrated in Fig.~\ref{xyz}, and $\bm{\Omega}$ denotes the helicity and decay parameters, with two distinct parametrization schemes for the helicity parameters.
For the first parametrization, the parameters $h_i$ and $\phi'_i$ ($i=1,3,4$)~\cite{BESIII:2020lkm} define the ratios of the helicity amplitudes and their relative phase differences, while in the second parametrization the helicity amplitudes are expressed in terms of the angular distribution parameters $\alpha_{\psi}$, $\alpha_1$, and $\alpha_2$, with $\phi_i$ ($i=1,3,4$)~\cite{Zhang:2023box} denoting the relative phase differences. Since the second parametrization corresponds to that for spin-1/2 particles, it provides a simpler connection to the polarization observables. This analysis is primarily based on the second parametrization, while the results obtained with the first parametrization are also provided. The correspondence between the two parametrizations is detailed in Ref.~\cite{Zhang:2023box}. Additionally, the weak decay parameters $\alpha_{\Xi^-}$, $\alpha_{\bar{\Xi}^+}$, $\phi_{\Xi^-}$, $\phi_{\bar{\Xi}^+}$, $\alpha_{\Lambda}$, and $\alpha_{\bar{\Lambda}}$ are well defined in Ref.~\cite{Perotti:2018wxm}.


\begin{figure*}[hbpt]
    \begin{center}
    \includegraphics[width=0.32\textwidth]{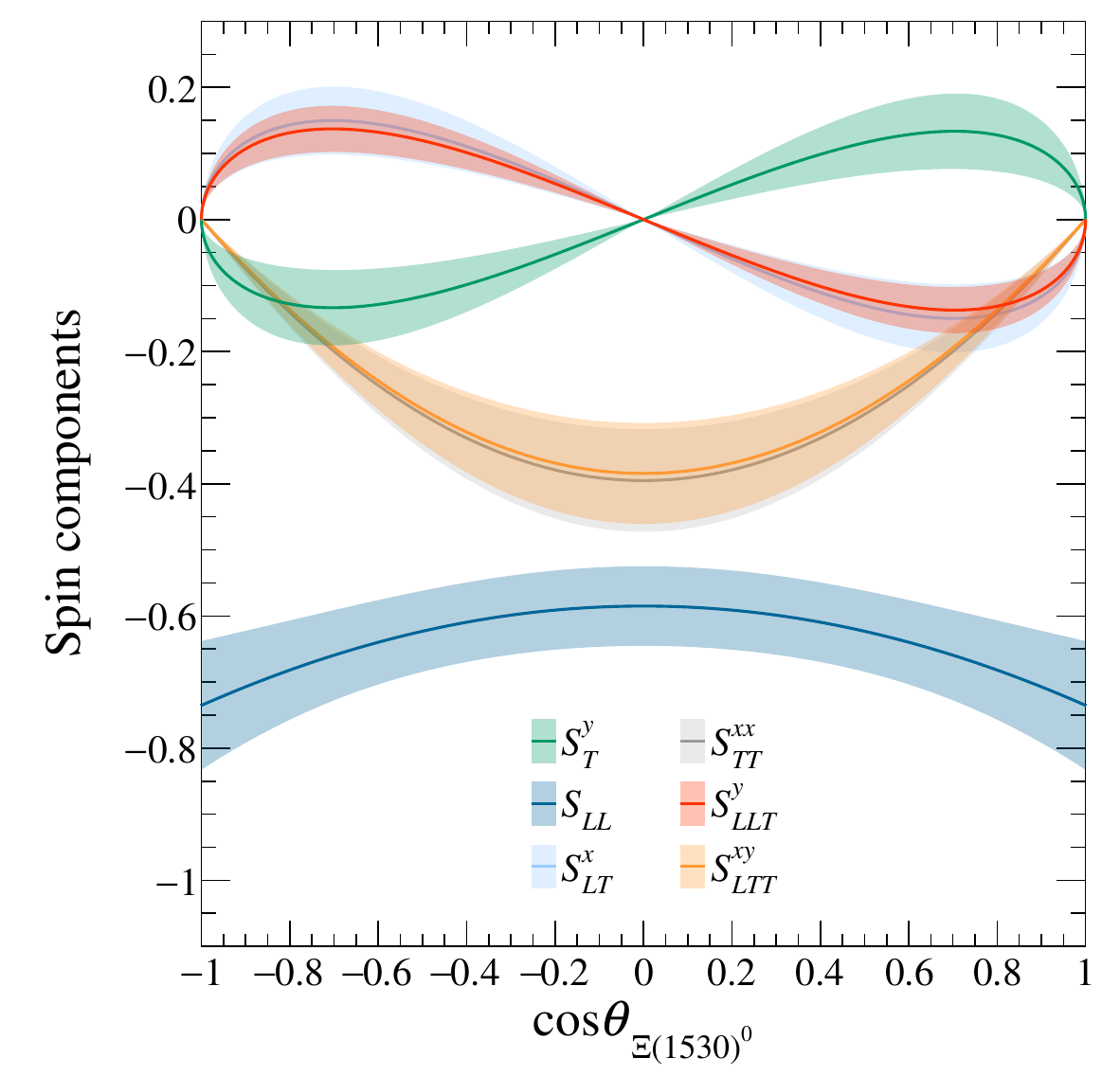}
    \includegraphics[width=0.32\textwidth]{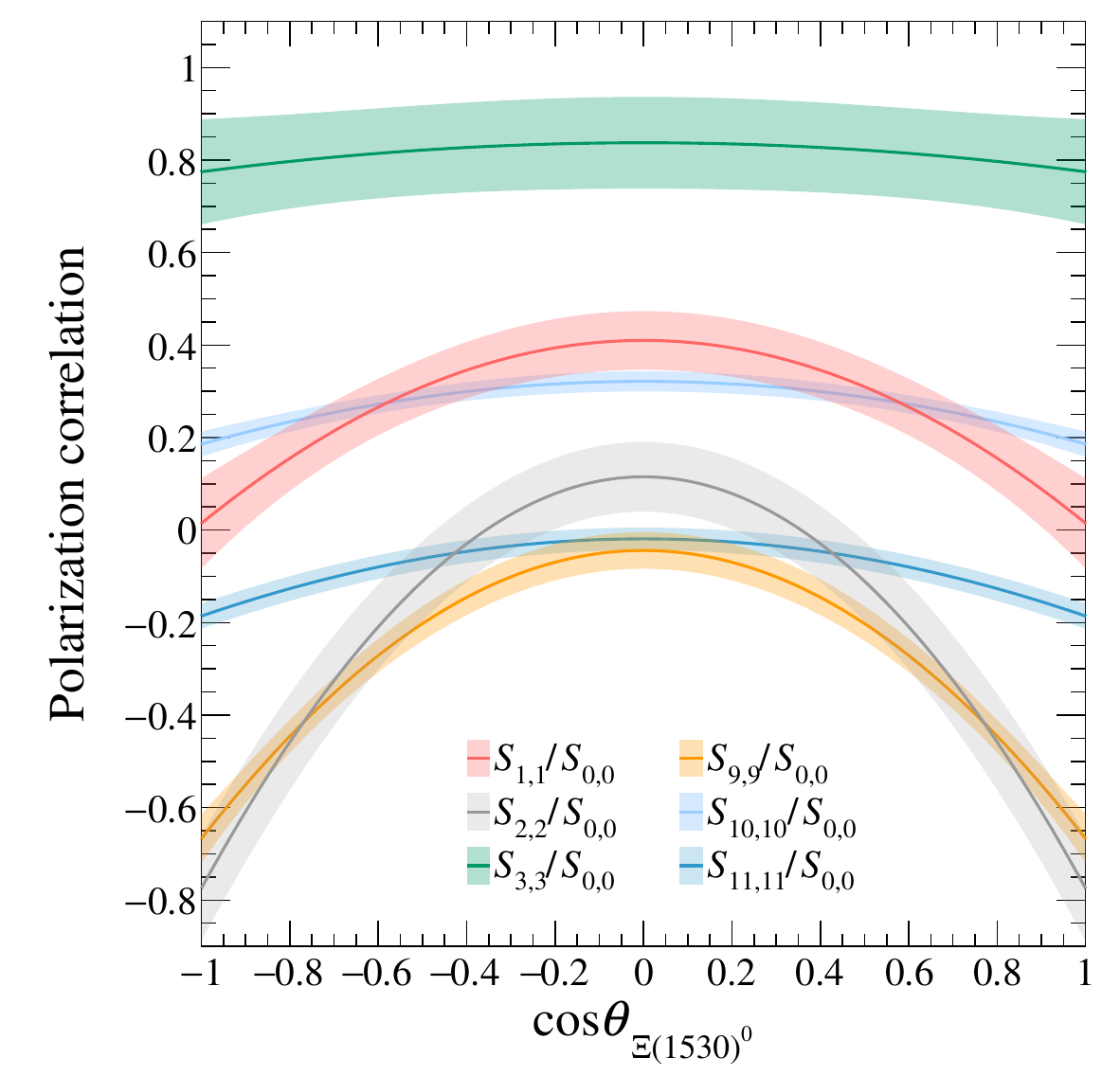}
    \includegraphics[width=0.32\textwidth]{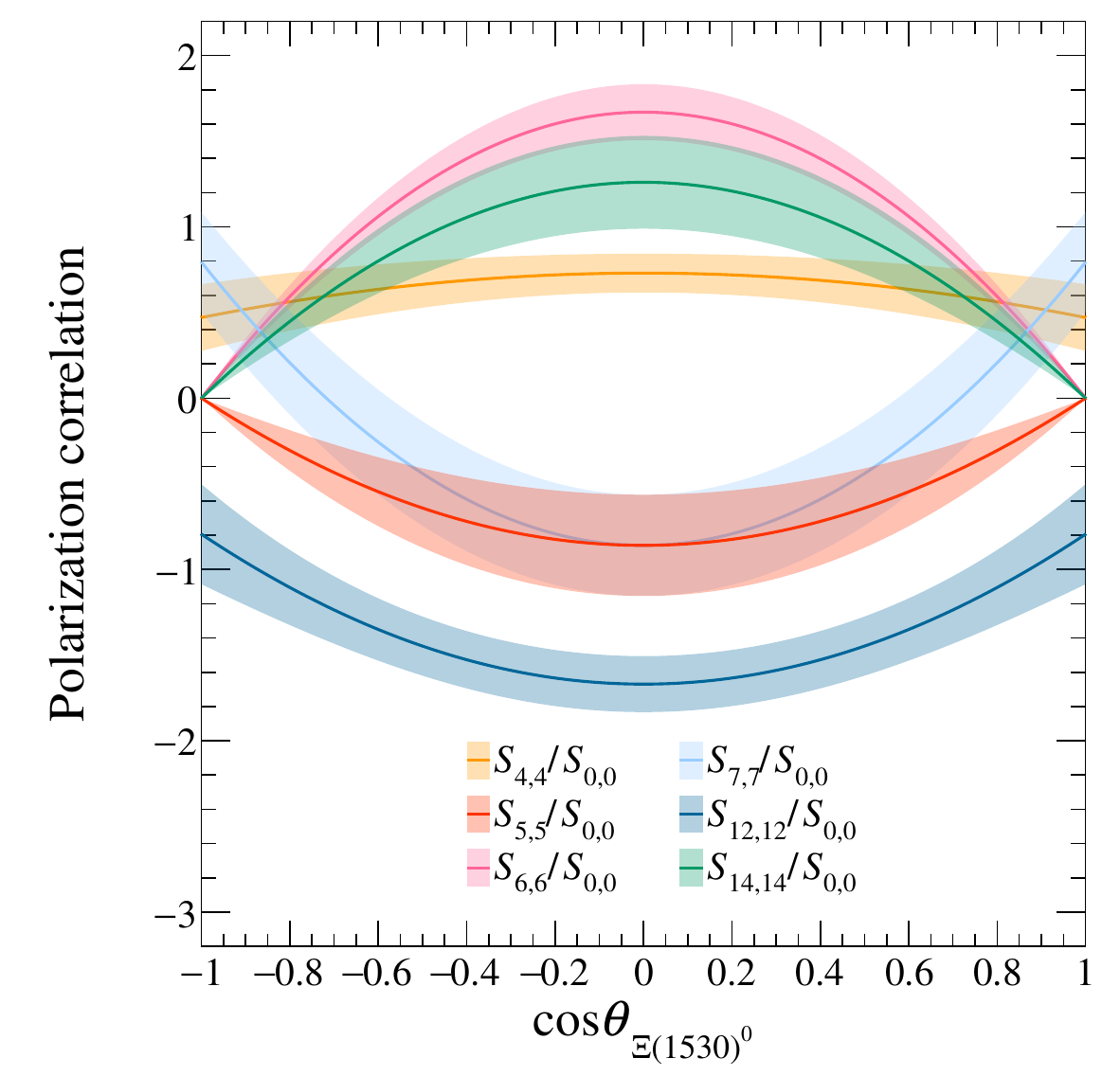}
    \end{center}
    \vspace{-0.5cm}
    \caption{The $\cos\theta_{\Xs}$ dependence of the nonzero independent polarization components (left) and the polarization correlation terms (middle and right). Here, $S_{i,i}$ ($i=0,1,2,...,14$) denotes the diagonal elements of the $S$ matrix, with $S_{0,0}$ representing the cross section term used for normalization. The solid lines represent the central values, and the shaded areas represent the ranges of 1 standard deviation.}
    \label{pol}
\end{figure*}

An unbinned maximum likelihood fit is performed to extract the set of $\Xs$ polarization parameters $\bm{\Omega}$. In the fit, the weak decay parameters for both $\Xi^-(\bar\Xi^+)$ and $\Lambda(\bar\Lambda)$ are fixed according to the current highest-precision measurements~\cite{BESIII:2021ypr,BESIII:2022qax}, assuming charge-parity conservation. The likelihood function $\mathscr{L}$ is constructed based on the probability density function, $\mathscr{P}(\bm{\xi}_i)$, for the event $i$ characterized by the measured angles $\bm{\xi}_i$
\begin{equation}
    \mathscr{L}=\prod_{i=1}^{N}\mathscr{P}(\bm{\xi}_i;\bm{\Omega})=\prod_{i=1}^{N}\mathcal{C}\mathscr{W}(\bm{\xi}_i;\bm{\Omega})\epsilon(\bm{\xi}_i),
\end{equation}
where $N$ is the number of events in the signal region, the joint angular distribution $\mathscr{W}(\bm{\xi}_i;\bm{\Omega})$ is given in Eq.~(\ref{cs}), and $\epsilon(\bm{\xi}_i)$ is the detection efficiency. 
The normalization factor $\mathcal{C}^{-1}=\frac{1}{N_{\rm MC}}\sum_{j=1}^{N_{\rm MC}}\mathscr{W}(\bm{\xi}_j;\bm{\Omega})$ is calculated as the sum of the corresponding amplitudes $\mathscr{W}$ from the accepted PHSP MC events $N_{\rm MC}$, applying the same event selection criteria as for the data. 
The minimization of the objective function defined as $\mathcal{S} = -\ln\mathscr{L}_{\rm S} + \ln\mathscr{L}_{\rm B}$ is conducted using the MINUIT package from the ROOT library~\cite{James:1975dr}. 
Here, $\mathscr{L}_{\rm S}$ and $\mathscr{L}_{\rm B}$ represent the likelihood functions for events chosen in the signal and sideband regions, respectively.
Since the signal and sideband regions cover different mass ranges and thus have different efficiencies, we apply a multidimensional, machine-learning-based correction to the sideband data to match the signal region efficiency, with correction factors derived from MC simulations.

\begin{figure}[!hbpt]
    \begin{center}
    \includegraphics[width=0.48\textwidth]{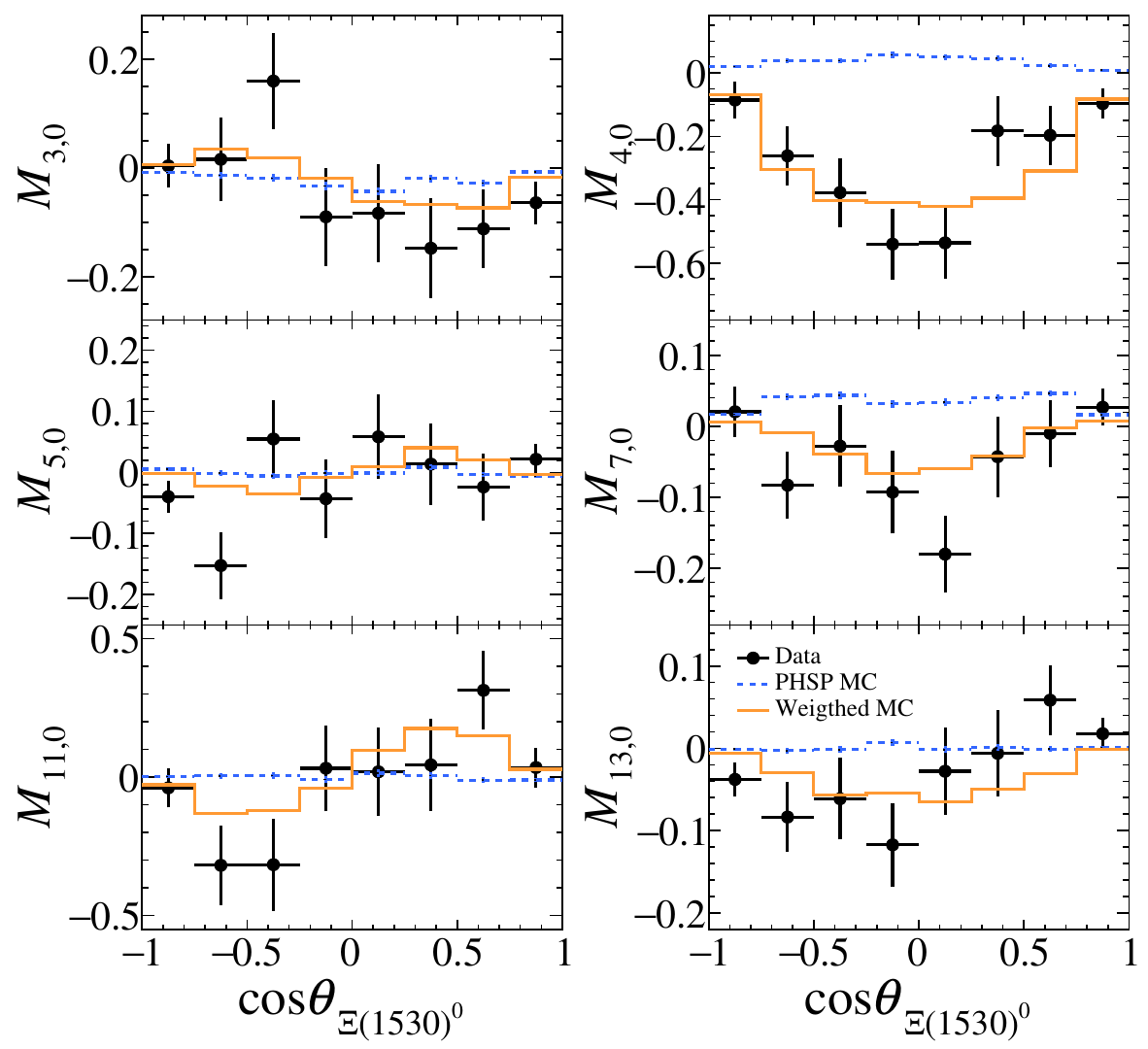}
    \end{center}
    \vspace{-0.5cm}
    \caption{Comparison of $M_{\mu\nu}$ distributions between data and MC. The blue histograms represent the PHSP MC samples, while the orange histograms represent the MC weighted by the measured parameters. The black dots with error bars are data.}
    \label{mom}
\end{figure}

The numerical results obtained from the maximum likelihood fit to the data, along with the moduli of the form factors and their ratios calculated based on the relationships provided in Ref.~\cite{Zhang:2023box}, are summarized in Table~\ref{tab:solution}.
The first row and the first column of the matrix $S$ in Eq.~(\ref{cs}) represent the spin polarization components of $\Xs$ and $\Xsb$, respectively, while its diagonal terms denote the polarization correlation coefficients between $\Xs$ and $\Xsb$. Among these, the nonzero independent terms, calculated based on the measured parameters, include the spin vector ($S_T^y$), rank-2 spin tensor ($S_{LL}$, $S_{LT}^x$, $S_{TT}^{xx}$), and rank-3 spin tensor ($S_{LLT}^y$, $S_{LTT}^{xy}$)~\cite{Zhang:2023wmd,Zhang:2023box} for the polarization components, as well as the independent terms for the correlation coefficients, all of which are displayed in Fig.~\ref{pol}.
To illustrate the model’s description of the data, we define the moments 
\begin{align}
    M_{\mu\nu}(\cos\theta_{\Xi(1530)^0}) =  &\frac{1}{N}\sum_{n=0}^{N}\sum_{i=0}^{3}\sum_{j=0}^{3}\sum_{k=0}^{3}\sum_{l=0}^{3}b^{\Xi(1530)^0}_{\mu i}b^{\bar\Xi(1530)^0}_{\nu j} \\ \notag
    &\times a^{\Xi^-}_{ik}a^{\bar\Xi^+}_{jl}a^{\Lambda}_{k0}a^{\bar\Lambda}_{l0}
\end{align}
for projecting six polarization terms of the $\Xi(1530)^0$, as shown in Fig.~\ref{mom}. Here, $N$ denotes the number of signal events in the data, and $\mu\nu$ represents the corresponding element of the matrix $S$ in Eq.~(\ref{cs}). 
For each polarization component, the polarization degree is evaluated by integrating its absolute value over $\theta_{\Xs}$, and the significance is determined by the ratio of the central value to its uncertainty. The significances of $S_T^y$, $S_{LL}$, $S_{LT}^x$, $S_{TT}^{xx}$, $S_{LLT}^{y}$, and $S_{LTT}^{xy}$, after taking into account systematic uncertainties, are $3.3\sigma$, $16.9\sigma$, $2.6\sigma$, $8.5\sigma$, $5.6\sigma$, and $8.1\sigma$, respectively. The total significance, obtained by weighting all components~\cite{Zhang:2023box}, is $20.9\sigma$.

\begin{table}[!htbp]
    \caption{Numerical results and uncertainties of the helicity parameters, the moduli of form factors, and their ratios. The first and second uncertainties are statistical and systematic, respectively.}
    \centering
    \scalebox{0.88}{
    \begin{tabular}{c c  c r}
    \hline
    \hline
    Parameter        &\multicolumn{1}{c}{Value}  &Parameter        &\multicolumn{1}{c}{Value}\\ 
    \hline 
    $\alpha_{\psi}$  &$~~0.02\pm0.06\pm0.06$      &$h_1$            &$ 0.69\pm0.05\pm0.05$ \\
    $\alpha_1$       &$-0.42\pm0.08\pm0.05$       &$h_3$            &$ 0.43\pm0.04\pm0.04$ \\    
    $\alpha_2$       &$~~0.75\pm0.06\pm0.06$      &$h_4$            &$ 0.43\pm0.03\pm0.03$ \\
    $\phi_1$ (rad)   &$-1.81\pm0.13\pm0.07$       &$\phi_1'$ (rad)  &$ 0.56\pm0.11\pm0.06$ \\
    $\phi_3$ (rad)   &$~~2.37\pm0.08\pm0.05$      &$\phi_3'$ (rad)  &$ 2.37\pm0.08\pm0.05$ \\
    $\phi_4$ (rad)   &$~~1.02\pm0.15\pm0.08$      &$\phi_4'$ (rad)  &$-2.90\pm0.13\pm0.08$ \\[1ex]
    $|G_E|$          &$0.0035\pm0.0006\pm0.0004$  &$|G_M/G_E|$      &$1.61\pm0.33\pm0.22$  \\
    $|G_M|$          &$0.0056\pm0.0006\pm0.0003$  &$|G_Q/G_E|$      &$3.71\pm0.63\pm0.38$  \\
    $|G_Q|$          &$0.0129\pm0.0007\pm0.0005$  &$|G_O/G_E|$      &$1.13\pm0.20\pm0.13$  \\
    $|G_O|$          &$0.0039\pm0.0002\pm0.0001$  &&                 \\
    \hline
    \hline
    \end{tabular}}
    \label{tab:solution}
\end{table}


\begin{table*}[!htbp]
    \caption{Absolute systematic uncertainties for the measured helicity parameters. The parameters $\phi_i$ and $\phi_i'$ ($i=1,3,4$) are given in radians.}
    \centering
    \scalebox{1.0}{
    \begin{tabular}{l c c c c c c | c c c c c c }
    \hline
    \hline
    Source                  &$\alpha_{\psi}$ &$\alpha_1$ &$\alpha_2$ &$\phi_1$ &$\phi_3$ &$\phi_4$ &$h_1$ &$h_3$ &$h_4$ &$\phi_1'$ &$\phi_3'$ &$\phi_4'$\\ 
    \hline          
    Reconstruction        ~~~~&~0.05~~ &~~0.04~~ &~~0.05~~ &~~0.03~~ &~~0.02~~ &~~0.05~~    &~~0.04~~ &~~0.03~~ &~~0.03~~ &~~0.03~~ &~~0.02~~ &~~0.05~~ \\
    Backgrounds           ~~~~&~0.01~~ &~~0.02~~ &~~0.01~~ &~~0.04~~ &~~0.04~~ &~~0.04~~    &~~0.02~~ &~~0.02~~ &~~0.01~~ &~~0.04~~ &~~0.04~~ &~~0.04~~ \\
    Decay parameters      ~~~~&~0.00~~ &~~0.00~~ &~~0.00~~ &~~0.01~~ &~~0.00~~ &~~0.00~~    &~~0.00~~ &~~0.00~~ &~~0.00~~ &~~0.00~~ &~~0.00~~ &~~0.00~~ \\
    Fit method            ~~~~&~0.02~~ &~~0.01~~ &~~0.01~~ &~~0.01~~ &~~0.00~~ &~~0.02~~    &~~0.01~~ &~~0.01~~ &~~0.01~~ &~~0.01~~ &~~0.00~~ &~~0.02~~ \\
    Efficiency correction ~~~~&~0.01~~ &~~0.01~~ &~~0.01~~ &~~0.01~~ &~~0.01~~ &~~0.02~~    &~~0.01~~ &~~0.00~~ &~~0.00~~ &~~0.01~~ &~~0.01~~ &~~0.01~~ \\
    Width                 ~~~~&~0.00~~ &~~0.01~~ &~~0.01~~ &~~0.04~~ &~~0.02~~ &~~0.04~~    &~~0.01~~ &~~0.01~~ &~~0.00~~ &~~0.04~~ &~~0.02~~ &~~0.04~~ \\[1ex]

    Total                   &0.06 &0.05 &0.06 &0.07 &0.05 &0.08                &0.05 &0.04 &0.03 &0.06 &0.05 &0.08             \\
    
    \hline
    \hline
    \end{tabular}}
    \label{tab:systematic}
\end{table*}

The systematic uncertainties in this analysis mainly originate from the reconstruction of $\Xs$ and $\Xsb$, the $1C$ kinematic fit, the backgrounds, the decay parameters, the fit method, the multidimensional MC efficiency correction, and the width of $\Xs$ and $\Xsb$.
The first part of the systematic uncertainties includes the $\Xi^-$ and $\bar\Xi^+$ reconstruction, the selection of the pion decaying from $\Xs$ or $\Xsb$, and the mass requirements for $\Xs$ and $\Xsb$. The uncertainties for the $\Xi^-$ and $\bar\Xi^+$ reconstruction and the pion track selection are cited from Refs.~\cite{BESIII:2022lsz,Liu:2024uno}, while the uncertainty due to the $\Xs$ and $\Xsb$ mass requirements is evaluated using the same method as described in Ref.~\cite{BESIII:2020fqg}.
The systematic uncertainty associated with the $1C$ kinematic fit is estimated by comparing results with and without track helix parameter corrections~\cite{BESIII:2024jgy}, and it is found to be negligible. 
The systematic uncertainties due to the backgrounds, including contributions from the sideband, continuum processes, and other nonpeaking backgrounds, are evaluated as follows: the systematic uncertainty from the sideband is assessed by varying the input background yields by 1 standard deviation, while the uncertainties from other backgrounds are estimated by comparing fits with and without their inclusion.
The uncertainties from decay parameters are evaluated by individually varying each fixed value by 1 standard deviation~\cite{BESIII:2021ypr,BESIII:2022qax}, and the sum of the largest differences is taken as the systematic uncertainty.
To estimate the systematic uncertainty of the fitting method, multiple sets of MC simulations generated with custom input parameters matching the data size are employed. The differences between the input and output values are taken as the systematic uncertainties.
For the multidimensional MC efficiency correction, to account for statistical fluctuations, multiple sets of correction factors are repeatedly extracted and applied to the data for efficiency correction and polarization fitting, and the standard deviation of the resulting distributions is taken as the systematic uncertainty.
Additionally, the width of $\Xs$ and $\Xsb$ is varied by 1 standard deviation~\cite{ParticleDataGroup:2024cfk} in the MC simulation, and the largest difference is taken as the systematic uncertainty. 
The absolute systematic uncertainties for various sources are summarized in Table~\ref{tab:systematic}. 
Assuming all sources are independent, the total systematic uncertainty is obtained by adding the individual contributions in quadrature.


In summary, based on $(2712.4\pm14.3)\times10^6$ $\psi(3686)$ events collected with the BESIII detector at the BEPCII collider, we report the first angular distribution analysis for a decuplet baryon excited state. 
For the process $\EE\to\psi(3686)\to\Xs\Xsb$, followed by the decays of both $\Xs$ and $\Xsb$, the spin structure is fully decomposed, and the helicity parameters are determined for the first time using two distinct parametrization schemes, as shown in Table~\ref{tab:solution}. 
Among these, the angular distribution parameter is measured to be $\alpha_{\psi} = 0.02\pm0.06\pm0.06$, which is consistent with the previous measurement~\cite{BESIII:2019dve, BESIII:2021gca} within $2\sigma$.
With these measured parameters, the $\cos\theta_{\Xs}$ dependence of the multipolar polarization components and correlations are calculated, as displayed in Fig.~\ref{pol}. 
This demonstrates that the $\Xs$ exhibits not only vector polarization ($S_T^y$), but also quadrupole polarization ($S_{LL}$, $S_{LT}^x$, $S_{TT}^{xx}$), and octupole polarization ($S_{LLT}^y$, $S_{LTT}^{xy}$). 
The significances of $S_T^y$, $S_{LL}$, $S_{LT}^x$, $S_{TT}^{xx}$, $S_{LLT}^{y}$, and $S_{LTT}^{xy}$, after accounting for systematic uncertainties, are $3.3\sigma$, $16.9\sigma$, $2.6\sigma$, $8.5\sigma$, $5.6\sigma$, and $8.1\sigma$, respectively. Combining all components with weights, the total significance is $20.9\sigma$.

Additionally, the moduli of the form factors and their ratios for $\Xs$ are determined for the first time, as summarized in Table~\ref{tab:solution}, with the nonzero electric quadrupole and magnetic octupole moments indicating a nonspherical charge distribution. Moreover, a comparison of the moduli ratios with those of $\Omega^-$~\cite{BESIII:2020lkm} reveals that $\Xs$ exhibits a more pronounced multipole moment effect. These measurements of higher-order multipole moments address the current scarcity of experimental data, offering valuable physical insights into baryon structure and serving as a critical reference for lattice QCD and quark model studies.

$Acknowledgments$---The BESIII Collaboration thanks the staff of BEPCII~\cite{BEPCII:link} and the IHEP computing center for their strong support. This work is supported in part by 
National Key R\&D Program of China under Contracts No. 2023YFA1606000 and No. 2023YFA1606704; National Natural Science Foundation of China (NSFC) under Contracts No. 12075107,No. 12247101, No. 12305085, No. 12247121, No. 11635010, No. 11935015, No. 11935016, No. 11935018, No. 12025502, No. 12035009, No. 12035013, No. 12061131003, No. 12192260, No. 12192261, No. 12192262, No. 12192263, No. 12192264, No. 12192265, No. 12221005, No. 12225509, No. 12235017, No. 12342502, and No. 12361141819; 
the Fundamental Research Funds for the Central Universities No. lzujbky-2025-ytA05, No. lzujbky-2025-it06, and No. lzujbky-2024-jdzx06;
the Natural Science Foundation of Gansu Province No. 22JR5RA389 and No.25JRRA799;
the ``111 Center'' under Grant No. B20063;
National Ten Thousand Talents Program for Top Young Talents, 
the Chinese Academy of Sciences (CAS) Large-Scale Scientific Facility Program; the Strategic Priority Research Program of Chinese Academy of Sciences under Contract No. XDA0480600; CAS under Contract No. YSBR-101; 100 Talents Program of CAS; The Institute of Nuclear and Particle Physics (INPAC) and Shanghai Key Laboratory for Particle Physics and Cosmology; ERC under Contract No. 758462; German Research Foundation DFG under Contract No. FOR5327; Istituto Nazionale di Fisica Nucleare, Italy; Knut and Alice Wallenberg Foundation under Contracts No. 2021.0174, No. 2021.0299, and No. 2023.0315; Ministry of Development of Turkey under Contract No. DPT2006K-120470; National Research Foundation of Korea under Contract No. NRF-2022R1A2C1092335; National Science and Technology fund of Mongolia; Polish National Science Centre under Contract No. 2024/53/B/ST2/00975; STFC (United Kingdom); Swedish Research Council under Contract No. 2019.04595; U. S. Department of Energy under Contract No. DE-FG02-05ER41374.

$Data~availability$---There are no publicly available research data or software supporting this manuscript. Requests for further information or data should be sent to the authors.

\bibliography{Draft_Xi1530_ref.bib}

\begin{widetext}
\begin{center}
\small
M.~Ablikim$^{1}$\BESIIIorcid{0000-0002-3935-619X},
M.~N.~Achasov$^{4,c}$\BESIIIorcid{0000-0002-9400-8622},
P.~Adlarson$^{81}$\BESIIIorcid{0000-0001-6280-3851},
X.~C.~Ai$^{86}$\BESIIIorcid{0000-0003-3856-2415},
C.~S.~Akondi$^{31A,31B}$\BESIIIorcid{0000-0001-6303-5217},
R.~Aliberti$^{39}$\BESIIIorcid{0000-0003-3500-4012},
A.~Amoroso$^{80A,80C}$\BESIIIorcid{0000-0002-3095-8610},
Q.~An$^{77,64,\dagger}$,
Y.~H.~An$^{86}$\BESIIIorcid{0009-0008-3419-0849},
Y.~Bai$^{62}$\BESIIIorcid{0000-0001-6593-5665}
% ...
\\
\vspace{0.2cm}
(BESIII Collaboration)\\
\vspace{0.2cm} {\it
$^{1}$ Institute of High Energy Physics, Beijing 100049, People's Republic of China\\
% ...
$^{4}$ Budker Institute of Nuclear Physics SB RAS (BINP), Novosibirsk 630090, Russia\\
% ...
$^{31}$ INFN Sezione di Ferrara, (A)INFN Sezione di Ferrara, I-44122, Ferrara, Italy; (B)University of Ferrara, I-44122, Ferrara, Italy\\
% ...
$^{39}$ Johannes Gutenberg University of Mainz, Johann-Joachim-Becher-Weg 45, D-55099 Mainz, Germany\\
% ...
$^{62}$ Southeast University, Nanjing 211100, People's Republic of China\\
% ...
$^{64}$ State Key Laboratory of Particle Detection and Electronics, Beijing 100049, Hefei 230026, People's Republic of China\\
% ...
$^{77}$ University of Science and Technology of China, Hefei 230026, People's Republic of China\\
% ...
$^{80}$ University of Turin and INFN, (A)University of Turin, I-10125, Turin, Italy; (B)University of Eastern Piedmont, I-15121, Alessandria, Italy; (C)INFN, I-10125, Turin, Italy\\
$^{81}$ Uppsala University, Box 516, SE-75120 Uppsala, Sweden\\
% ...
$^{86}$ Zhengzhou University, Zhengzhou 450001, People's Republic of China\\

\vspace{0.2cm}
$^{\dagger}$ Deceased\\
% ...
$^{c}$ Also at the Novosibirsk State University, Novosibirsk, 630090, Russia\\
% ...
}
\end{center}
\end{widetext}

\end{document}